\documentclass[a4paper]{jpconf}
\usepackage{graphicx}
\usepackage{amssymb}
\usepackage{amsmath}
\usepackage{dsfont}

\newcommand{\R}{\mathds{R}}

\newcommand{\Acal}{{\cal A}}
\newcommand{\Fcal}{{\cal F}}

\newcommand{\Ical}{{\cal I}}

\newcommand{\Mcal}{{\cal M}}

\def\im{\mathrm{i}}
\def\ep{\mathrm{e}}
\def\pa{\partial}
\def\diff{\mathrm{d}}

\def\sfrac#1#2{{\textstyle\frac{#1}{#2}}}
\def\]{\right]}
\def\[{\left[}
\def\){\right)}
\def\({\left(}
\def\>{\rangle}
\def\<{\langle}
\def\+{\dagger}

\def\={\ =\ }
\def\und{\quad\textrm{and}\quad}
\def\with{\quad\textrm{with}\quad}

\begin{document}
\title{On Yang--Mills fields from anti-de Sitter spaces}

\author{Kaushlendra Kumar}

\address{Institute of Physics, Humboldt University, Zum Großen Windkanal 2, 12489 Berlin}

\ead{kaushlendra.kumar@hu-berlin.de}

\begin{abstract}
Motivated by some recent progress involving a non-compact gauge group, we obtain classical gauge fields using non-compact foliations of anti-de Sitter space in $4$ dimensions (required dimensionality for conformal invariance of Yang--Mills theory) and transfer these to Minkowski spacetime using a series of conformal maps. This construction also builds upon some previous works involving $\textrm{SU}(2)$ gauge group in that we now use its non-compact counterpart $\textrm{SU}(1,1)$ here. We note down gauge fields in both Abelian as well as non-Abelian settings and find them to be divergent at some hyperboloid, which is a hypersurface of co-dimension $1$ inside the conformal boundary of $\textrm{AdS}_4$. In spite of this hurdle we find a physically relevant field configuration in the Abelian case, reproducing a known result.
\end{abstract}

\section{Introduction}
Vacuum Yang--Mills theory in $4$ dimensions is conformally invariant---a fact that can be used to map their solutions to any conformally related spacetime\footnote{It is a known fact that all three variants of FLRW spacetimes viz.~Minkowski $\R^{1,3}$, de Sitter $\textrm{dS}_4$ and anti-de Sitter $\textrm{AdS}_4$ are covered this way.} after obtaining them on a more suitable one equipped with symmetric-foliations, and therefore admitting a natural gauge group. This exercise was carried out for $\textrm{dS}_4$ whose foliation submanifold $S^3$ arises as group manifold of $\textrm{SU}(2)$; the resulting $\textrm{SU}(2)$-equivariant connection ansatz reproduced \cite{zhilin} a well known Yang--Mills solutions obtained in 1977 by L\"uscher \cite{luescher}. This was achieved by following set of conformal correspondences: $\Ical\times S^3 \xleftarrow{\text{conformal}} \textrm{dS}_4  \xrightarrow{\text{conformal}} \R^{1,3}$ where the cylinder domain $\Ical =(-\sfrac\pi2,\sfrac\pi2)$ needs to be doubled to facilitate a gluing of two $\textrm{dS}_4$ copies so as to cover entire Minkowski spacetime. Furthermore, solutions for its Abelian counterpart $U(1)\subset \textrm{SU}(2)$ (in a non-symmetric setting) produces a family of electromagnetic knotted fields \cite{zhilin,kumar}. These EM knots have rather intriguing physical features, many of which have been explored recently \cite{hantzko,KLP22}. Another motivation for such an exploration of classical gauge fields is that only a few of them are known in mathematical physics literature (see e.g.~\cite{actor,rajaraman,manton} for reviews). To this end, novel results were obtained on de Sitter and anti-de Sitter spaces in \cite{IvLePo1,IvLePo2} for $\textrm{SU}(2)$ as well as some higher-dimensional generalizations (in de Sitter case) for $\textrm{SO}(4)$ in \cite{uenal}.

There has been some recent progress \cite{roehrig} towards obtaining gauge fields via $H^3$ and $\textrm{dS}_3$ foliations of Minkowski space regions with non-compact gauge group $\textrm{SO}(1,3)$, as opposed to compact ones discussed above. Inspired by this success, we considered $\textrm{AdS}_3$ foliations of $\textrm{AdS}_4$ since the former arises as group manifold of $\textrm{SU}(1,1)$. Classical gauge fields were obtained here for $\textrm{SU}(1,1)$-symmetric gauge configuration \cite{Hir23} (of both Abelian and non-Abelian type) by employing following strategy: $\Ical\times \textrm{AdS}_3 \xleftarrow{\text{conformal}} \textrm{AdS}_4  \xrightarrow{\text{conformal}} \Ical\times S^3_+$ followed by the previous $\Ical\times S^3 \xrightarrow{\text{conformal}} \R^{1,3}$ after gluing two $\textrm{AdS}_4$ copies to recover full $S^3$. However, unlike the previous compact case, this gluing is not smooth and leads to a singular conformal boundary; this feature propagates to the corresponding gauge fields but in a milder fashion. We present such field configurations in a compact form and compute their stress energy tensor. We also make use of several plots to explain various features of these conformal maps and resulting field configurations (such as orientation preserving gluing).

\section{Geometrical toolkit for anti-de Sitter space $\mathrm{AdS}_{4}$}
We start with the following isometric embedding of $\mathrm{AdS}_{4}$---endowed with coordinates $(x^1,x^2,x^3,x^4,x^5)$ and global radius $R$---inside $\R^{2,3}$ via
\begin{align}\label{eqAdS4}
    -(x^{1})^2-(x^{2})^2+(x^{3})^2+(x^{4})^2+(x^{5})^2\=-R^2\ .
\end{align}

\subsection{$\mathrm{AdS}_{4}$-foliations of the form ${\cal I}\times \mathcal{M}_{3}$}
First, we have $\mathcal{M}_{3}=\mathrm{AdS}_{3}$ foliation with a spacelike parameter $\psi\in \Ical=(-\pi/2,\pi/2)$. This $\mathrm{AdS}_{3}\hookrightarrow\R^{2,2}$ can be described using embedding coordinates $\alpha^i(\rho,\tau,\phi)$ for $i=1,\ldots,4$ with spatial parameters $\rho\in\R_+$, $\phi\in [0,2\pi]$ and temporal parameter $\tau\in [0,2\pi]$ as
\begin{equation}
\begin{aligned}
    \alpha^{1} &= \cosh{\rho}\,\cos{\tau},\ \alpha^{2} = \cosh{\rho}\,\sin{\tau},\\
    \alpha^{3} &= \sinh{\rho}\,\cos{\phi},\ \alpha^{4}
    = \sinh{\rho}\,\sin{\phi},
\end{aligned}
\quad
\implies 
\quad
-(\alpha^1)^2-(\alpha^2)^2+(\alpha^3)^2+(\alpha^4)^2=-1\ .
\end{equation}
The following global $\textrm{AdS}_4$ embedding coordinates, 
\begin{equation}\label{coord1}
    x^i \= R\sec\chi\,\alpha^i,\qquad x^5 \= R\tan\chi\ ,
\end{equation}
then yields its induced metric (arising from flat $\R^{2,3}$ metric) in terms of $\textrm{AdS}_3$ metric $\diff\Omega^{2}_{1,2}$ as
\begin{align}\label{metric1}
    \mathrm{d}s^{2}\=\frac{R^{2}}{\cos^{2}\!{\psi}}
    \bigl(\diff\psi^2-\cosh^2\!\rho\,\diff\tau^2+\diff\rho^2+\sinh^2\!\rho\,\diff\phi^2\bigr) \=
    \frac{R^{2}}{\cos^{2}\!{\psi}}\bigl(\mathrm{d}\psi^{2}+\mathrm{d}\Omega^{2}_{1,2}\bigr)\ .
\end{align}

Next, we have $\Mcal_3{=}S^3_+$ (upper hemisphere) foliations, embedded inside $\R^{2,3}$ using
\begin{align}
    x^{1}=R\sec\chi\cos\tau,\
    x^{2}=R\sec\chi\sin\tau,\
    x^{3}=R\tan\chi\,\beta^1,\
    x^{4}=R\tan\chi\,\beta^2,\
    x^{5}=R\tan\chi\,\beta^3\ ,
\end{align}
where $\chi\in[0,\pi/2)$ for the half-sphere and canonical $S^2$ coordinates $\beta^i(\theta,\phi)$ with $\theta\in[0,\pi]$ are
\begin{align}
\beta^{1}\=\sin\theta\,\cos\phi\ ,\qquad\beta^{2}\=\sin\theta\,\sin\phi\ ,\qquad\beta^{3}\=\cos\theta\ . 
\end{align}
These are related to the coordinates ($\rho,\psi$) above \eqref{coord1} as follows,
\begin{equation}
    \tanh\rho \= \sin\theta\,\sin\chi \quad\und\quad \tan\psi \= -\cos\theta\,\tan\chi \ .
\end{equation}
In this case, the induced metric demonstrates a $S^{3}_{+}$-cylinder structure with round metric $\mathrm{d}\Omega^{2}_{3+}$; the latter can be expressed using $S^2$ round metric $\mathrm{d}\Omega^{2}_{2}$ as
\begin{align} \label{metric3}
    \mathrm{d}s^{2}\=\frac{R^{2}}{\cos^{2}\!{\chi}}\bigl(-\mathrm{d}{\tau}^{2}+\mathrm{d}\chi^{2}+\sin^{2}\!\chi\,\mathrm{d}\Omega^{2}_{2}\bigr)\=\frac{R^{2}}{\cos^{2}\!{\chi}}\bigl(-\mathrm{d}{\tau}^{2}+\mathrm{d}\Omega^{2}_{3+}\bigr)\ .
\end{align}
This temporal parameter $\tau$ can be extended to full $\R$ by going to the universal cover $\widetilde{\textrm{AdS}_4}$.

\subsection{Gluing of two $\mathrm{AdS}_{4}$ copies}
\begin{figure}
    \centering
    \begin{minipage}{.3\textwidth}
        \centering
        \includegraphics[width=\linewidth, height=0.15\textheight]{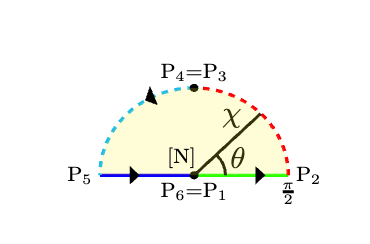}
        \caption{Depiction of $S^3_+$-boundary with colored segments between key points.}
        \label{glueLeft}
    \end{minipage}
    \hspace{2mm}
    \begin{minipage}{0.3\textwidth}
        \centering
        \includegraphics[width=\linewidth, height=0.15\textheight]{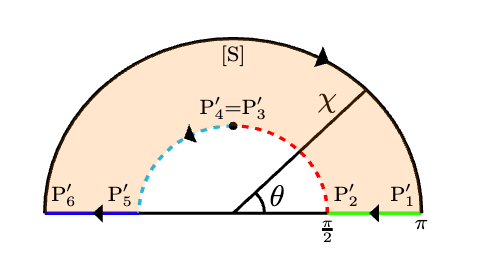}
        \caption{Depiction of $S^3_-$-boundary with colored segments between key points.}
        \label{glueCenter}
    \end{minipage}
    \hspace{2mm}
    \begin{minipage}{0.3\textwidth}
        \centering
        \includegraphics[width=\linewidth, height=0.15\textheight]{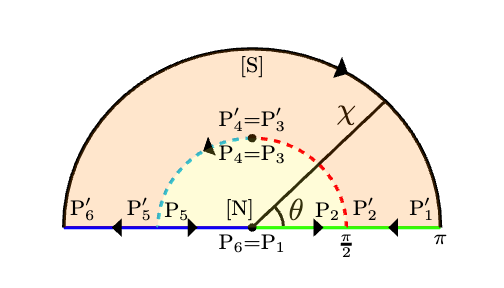}
        \caption{Depiction of the gluing across  $S^2$ boundary while preserving orientation.}
        \label{glueRight}
    \end{minipage}
\end{figure}

We now proceed to glue two copies of $\mathrm{AdS}_{4}$ to recover full round $3$-sphere, i.e.~$S^{3}=S^{3}_{+}\cup S^{2}\cup S^{3}_{-}$ in order to apply below-mentioned conformal map. To this end, we note down the following map that glues northern copy $S^3_+$ with southern one $S^3_-$ along the boundary $S^2$ at $\chi{=}\frac{\pi}{2}$:
\begin{equation}\label{gluing}
\begin{aligned}
    &\tanh{\rho} \= \varepsilon\,\sin{\theta}\,\sin{\chi}
    \und
    \tan{\psi} \= -\varepsilon\,\cos{\theta}\,\tan{\chi}\ ,\with\\
    \varepsilon|_{S^3_+}&=+1:\ \rho\in\R_+\Leftrightarrow \chi\in[0,\tfrac{\pi}{2}) \und
    \varepsilon|_{S^3_-}=-1:\ \rho\in\R_-\Leftrightarrow \chi\in(\tfrac{\pi}{2},\pi]\ .
\end{aligned}
\end{equation}
This map preserves the orientation along the gluing boundary $\partial S^3_\pm{=}S^2$. To see this, we note down some key points in table \ref{table} for both coordinate systems and then plot these in spherical coordinates $(\chi,\theta)$\footnote{This could also be demonstrated in other coordinate system as well; see \cite{Hir23} for details.} individually in figures \ref{glueLeft} and \ref{glueCenter} as well as in combined fashion in figure \ref{glueRight}. Notice in the figures that the open boundary of the AdS space (for fixed $\tau$) is depicted with dashed lines while some points (like the north pole) are identified as is clear from table \ref{table}. Finally, the gluing happens by identifying the points $(P_2,\ldots,P_5)$ with $(P_2',\ldots,P_5')$ pairwise so that same-colored segments are coincide and the orientation remains preserved. Another point to note here is that this gluing is not smooth, but consists of a singularity arising from the conformal boundary $\chi=\pi/2$ as is clear from above metric \eqref{metric3}.
\begin{table}
    \caption{\label{table} Key points on northern hemisphere $(\varepsilon{=}+1; P)$ and southern hemisphere $(\varepsilon{=}-1; P')$ in two coordinate systems for some infinitesimal $\epsilon>0$.}
    \begin{center}
    \begin{tabular}{llllll}
    \br
    $S^3_+$ & $(\rho,\psi)_+$ & $(\chi,\theta)_-$ & $S^3_-$ & $(\rho,\psi)_-$ & $(\chi,\theta)_-$\\
    \mr
    $\mathrm{P}_{1}$ & $(0,0{-}\epsilon)$ & $(0,0)$ & {$\mathrm{P}'_1$} & $(0,0{-}\epsilon)$ & $(\pi,0)$\\
    $\mathrm{P}_{2}$ & $(0,-\frac{\pi}{2})$ & $(\frac{\pi}{2},0)$ & {$\mathrm{P}'_2$} & $(0,-\frac{\pi}{2})$ & $(\frac{\pi}{2},0)$\\
    $\mathrm{P}_{3}$ & $(\infty,-\frac{\pi}{2})$ & $(\frac{\pi}{2},\frac{\pi}{2}{-}\epsilon)$ & {$\mathrm{P}'_3$} & $(-\infty,-\frac{\pi}{2})$ & $(\frac{\pi}{2},\frac{\pi}{2}{-}\epsilon)$\\
    $\mathrm{P}_{4}$ & $(\infty,\frac{\pi}{2})$ & $(\frac{\pi}{2},\frac{\pi}{2}{+}\epsilon)$ & {$\mathrm{P}'_4$} & $(-\infty,\frac{\pi}{2})$ & $(\frac{\pi}{2},\frac{\pi}{2}{+}\epsilon)$\\
    $\mathrm{P}_{5}$ & $(0,\frac{\pi}{2})$ & $(\frac{\pi}{2},\pi)$ & {$\mathrm{P}'_5$} & $(0,\frac{\pi}{2})$ & $(\frac{\pi}{2},\pi)$\\
    $\mathrm{P}_{6}$ & $(0,0{+}\epsilon)$ & $(0,\pi)$ & {$\mathrm{P}'_6$} & $(0,0{+}\epsilon)$ & $(\pi,\pi)$\\
    \br
\end{tabular}
\end{center}
\end{table}

\subsection{Conformal mapping to Minkowski space $\R^{1,3}$}
\begin{figure}[!htb]
    \centering
    \begin{minipage}{0.4\textwidth}
        \centering
        \includegraphics[width=\linewidth, height=0.3\textheight]{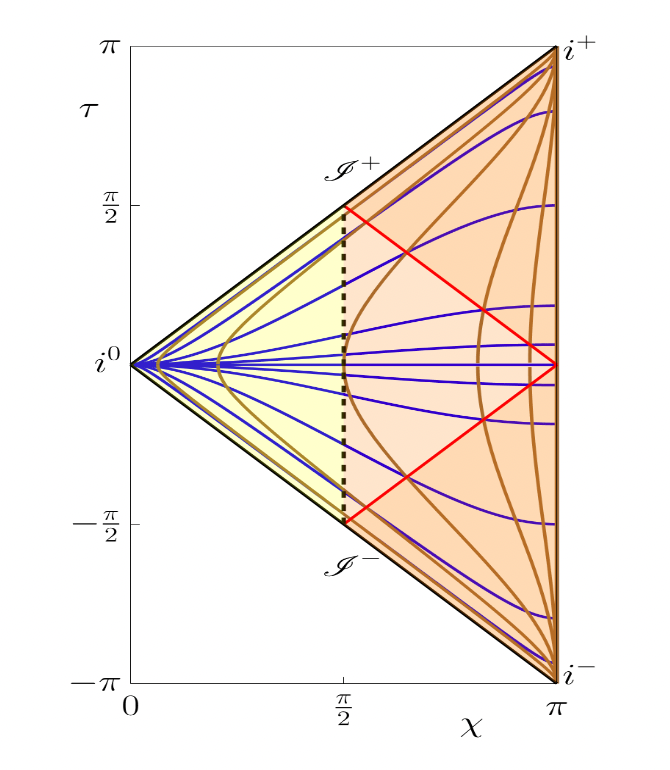}
        \caption{Penrose diagram inside AdS$_4$ spaces joined together (dotted black line) with constant (blue lines) $t$- and (brown lines) $r$-hypersurfaces and the lightcone at the origin (red lines).}
        \label{Mink1}
    \end{minipage}
    \hspace{5mm}
    \begin{minipage}{0.4\textwidth}
        \centering
        \includegraphics[width=\linewidth, height=0.3\textheight]{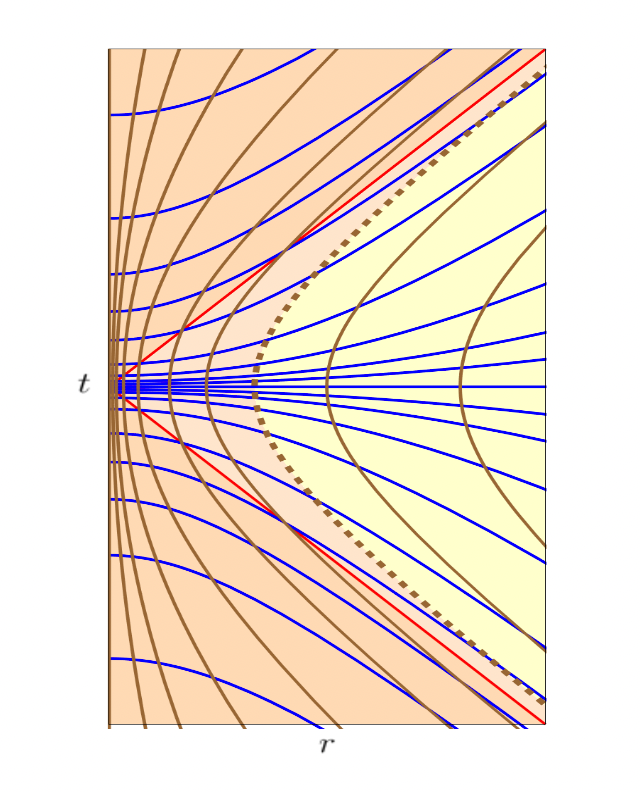}
        \caption{Minkwski spacetime in polar coordinates $(t,r)$ with constant $\tau$- \& $\chi$-slices shown in blue and brown lines respectively while the broken brown line shows the gluing-boundary.}
        \label{Mink2}
    \end{minipage}
\end{figure}

So far, we have seen an effective comformal map between $\textrm{AdS}_3$-cylinder \eqref{metric1} and $S^3$-cylinder (\ref{metric3},\ref{gluing}): $(\rho,\psi)\rightarrow(\chi,\theta)$ while keeping $\tau$ and $\phi$ fixed. We can now map the $S^3$-cylinder (post-gluing) to Minkowski space $\R^{1,3}$ equipped with polar coordinates $(t,r,\theta,\phi)$\footnote{Let us recall Minkowski polar coordinates: $(t,x,y,z)=(t,r\sin\theta\cos\phi,r\sin\theta\sin\phi,r\cos\theta)$.}: $(\tau,\chi)\rightarrow(t,r)$ via
\begin{align} \label{Mmap2}
    \sin\tau\=\gamma\frac{t}{R}
    \und
    \sin\chi\=\gamma\frac{r}{R}
    \with \gamma\=\frac{2R^{2}}{\sqrt{4R^{2}t^{2}+(r^{2}-t^{2}+R^{2})^{2}}}\ ,
\end{align}
where the $S^2$ coordinates $(\theta,\phi)$ are identified. A key features of this map is that the full Minkowski space gets embedded inside half of the doubled AdS domain with null-boundary given by $\chi{=}|\tau|$; this is due to the following inequality:
\begin{align} \label{chitau}
    \gamma\=\cos{\tau}-\cos{\chi}\ >0
    \qquad\implies\qquad\chi>\lvert\tau\rvert\ .
\end{align}
This feature is clearly exemplified through $(\tau,\chi)$ Penrose diagram in figure \ref{Mink1} demonstrating the flat spacetime embedding inside half of the (glued) $\textrm{AdS}_4$ space with future and past null boundaries. The $(t,r)$ plot in figure \ref{Mink2} further illustrates the gluing-boundary inside Minkowski spacetime. It should be noted here that every point inside shaded regions of these plots hides a $2$-sphere and that such regions coming from different $\textrm{AdS}_4$ copies are color-coded (yellow and orange shades).

The conformal metric \eqref{metric3} in these Minkowski coordinates take the following form:
\begin{equation} \label{metric4}
    \diff s^2 \= \frac{4\,R^4}{(r^2{-}t^2{-}R^2)^2}\,\bigl(-\diff t^2+\diff r^2+r^2\diff\Omega_2^2\bigr)\ ,
\end{equation}
which shows that the singularity at the gluing-boundary is a one-sheeted hyperboloid $H^{1,2}_R$:
\begin{equation}\label{hyperbola}
    \bigl\{\chi{=}\tfrac{\pi}{2}\bigr\} \qquad\Longleftrightarrow\qquad
    \bigl\{r^2{-}t^2{=}R^2\bigr\} \ =:\ H^{1,2}_R\ .
\end{equation}

\section{$\textrm{SU}(1,1)$-symmetric gauge theory on $\textrm{AdS}_4$}
Let us review some algebraic results required to construct the relevant connection one-form. To this end, we start with the group manifold of $\textrm{SU}(1,1)$ which is $\textrm{AdS}_3$, easily seen by the map
\begin{align}
    g:\;\mathrm{AdS_{3}}\, \rightarrow\,\mathrm{SU(1,1)} \qquad\mathrm{via}\qquad
    (\alpha^1,\alpha^2,\alpha^3,\alpha^4) \,\mapsto\, 
    \begin{pmatrix} \alpha^1{-}\im\alpha^2 & \alpha^3{-}\im\alpha^4 \\
    \alpha^3{+}\im\alpha^4 & \alpha^1{+}\im\alpha^2 \end{pmatrix}\ .
\end{align}
We use this $g$ to obtain left-invariant one-forms $e^{\alpha},\ \alpha=0,1,2$ via Maurer--Cartan method:
\begin{align}\label{MC1forms}
    \Omega_{L}(g)\=g^{-1}\mathrm{d}g \= e^{\alpha}\,I_{\alpha}
\end{align}
where $I_\alpha$ are $\mathfrak{sl}(2,\R)$ generators satisfying
\begin{align}
    [I_{\alpha},I_{\beta}] \= 2\,f_{\ \alpha\beta}^{\gamma}\,I_{\gamma} \quad\und\quad \mathrm{tr}(I_{\alpha}\,I_{\beta}) \= 2\,\eta_{\alpha\beta}\ ,
\end{align}
with $f^2_{\ 01}=f^1_{\ 20}=-f^0_{\ 12}=1$ and $(\eta_{\alpha\beta})=\mathrm{diag}(-1,1,1)$. The resulting one-forms look like,
\begin{equation}\label{1formsAdS}
\begin{aligned}
    e^{0}&\=\cosh^{2}\!{\rho}\;\mathrm{d}\tau+\sinh^{2}\!{\rho}\;\mathrm{d}\phi\ ,\\
    e^{1}&\=\cos{(\tau{-}\phi)}\;\mathrm{d}\rho+\sinh{\rho}\;\cosh{\rho}\;\sin{(\tau{-}\phi)}\;\diff{(\tau{+}\phi)}\ ,\\
    e^{2}&\=-\sin{(\tau{-}\phi)}\;\mathrm{d}\rho+\sinh{\rho}\;\cosh{\rho}\;\cos{(\tau{-}\phi)}\;\diff{(\tau{+}\phi)}\ .
\end{aligned}
\end{equation} 
These obey Cartan structure equation and provide orthonormal-frame on $\textrm{AdS}_3$-cylinder \eqref{metric1}:
\begin{equation}
    \diff e^{\alpha}+f^{\alpha}_{\ \beta\gamma}\;e^{\beta}\wedge e^{\gamma}=0 \quad\und\quad \diff s^{2}_{\mathrm{cyl}} \= \mathrm{d}\psi^{2}+\eta_{\alpha\beta}\,e^{\alpha}e^{\beta}\ .
\end{equation}

\begin{figure}
    \centering
    \begin{minipage}{.25\textwidth}
        \centering
        \includegraphics[width=\linewidth, height=0.2\textheight]{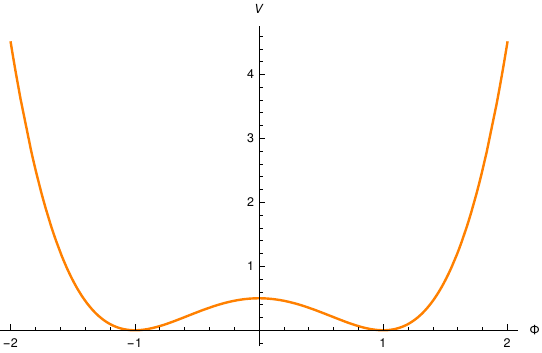}
        \caption{Plot of the potential $V(\Phi)$.}
        \label{potential}
    \end{minipage}
    \hspace{5mm}
    \begin{minipage}{0.32\textwidth}
        \centering
        \includegraphics[width=\linewidth, height=0.2\textheight]{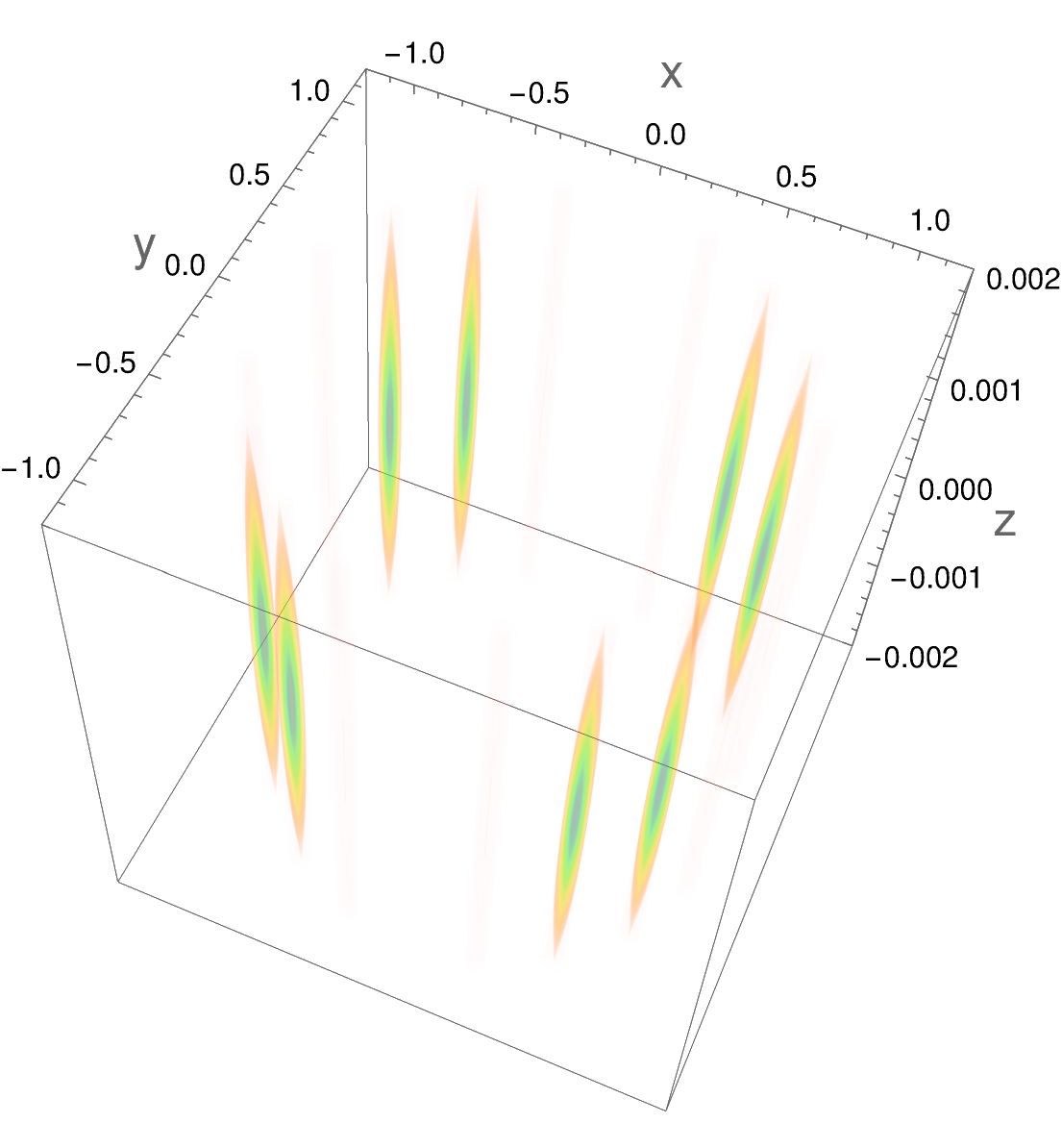}
        \caption{Color-graded density plot for the energy density $\rho\propto\frac{1}{(\lambda^2{+}4z^2)^2}|_{t=0}$ emphasizing its maxima near $z{=}0$.}
        \label{EnDenPlot}
    \end{minipage}
    \hspace{5mm}
    \begin{minipage}{0.32\textwidth}
        \centering
        \includegraphics[width=\linewidth, height=0.2\textheight]{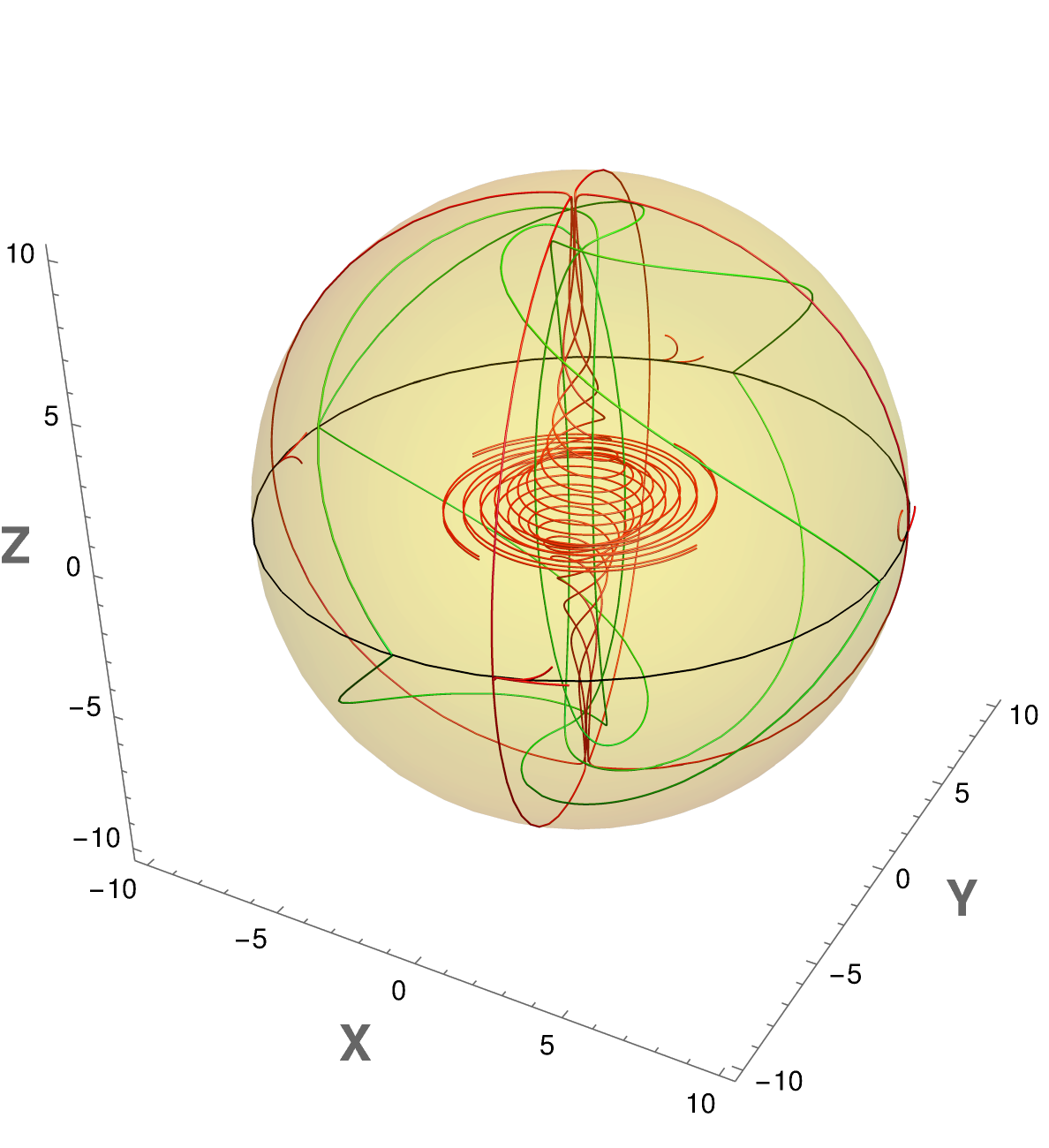}
        \caption{Electric (red) and magnetic (green) field lines for $\widetilde{{\bf S}}^{(0)}$ at $t{=}10$ and $\psi_0{=}\sfrac{\pi}{2}$ inside the sphere of radius $r{=}\sqrt{101}$.}
        \label{fieldLines}
    \end{minipage}
\end{figure}

Now a generic {\it gauge field} $\Acal$ in this frame can be made $\text{SU}(1,1)$-symmetric by,
\begin{equation}\label{gaugeField}
\begin{aligned}
    &\Acal \= \mathcal{A}_{\psi}\,e^{\psi}+ \mathcal{A}_{\alpha}\,e^{\alpha}\qquad\xrightarrow{\Acal_\psi=0}\qquad \mathcal{A}\=X_{\alpha}(\psi)\;e^{\alpha}\\
    &\implies \Fcal \= \diff\Acal + \Acal\wedge\Acal \= X_{\alpha}^{'}\;e^{\psi}\wedge e^{\alpha}+\tfrac{1}{2}\bigl(-2f_{\ \beta\gamma}^{\alpha}X_{\alpha}+[X_{\beta},X_{\gamma}]\bigr)\;e^{\beta}\wedge e^{\gamma}\ ,
\end{aligned}
\end{equation}
where $X'_\alpha$ in the field strength expression correspond to $\diff X_\alpha/\diff\psi$. Next, we impose the Gauss-law constraint $[X_{\alpha},X^{'}_{\alpha}]=0$ arising from the eom $*\diff*\Fcal = 0$ by following choice of components,
\begin{align} \label{Thetaansatz}
    X_{0}=\Theta_{0}(\psi)\,{I_{0}}\ ,\quad 
    X_{1}=\Theta_{1}(\psi)\,{I_{1}}\und
    X_{2}=\Theta_{2}(\psi)\,{I_{2}}\ .
\end{align}
The Yang--Mills Langrangian $\mathcal{L}=\sfrac14\mathrm{tr}(\Fcal\wedge*\Fcal)$ can then be readily computed to get,
\begin{equation} \label{matrixlag}
\begin{aligned}
    \mathcal{L} &\= \tfrac14\mathrm{tr}\mathcal{F}_{\psi\alpha}\mathcal{F}^{\psi\alpha}+\tfrac18\mathrm{tr}\mathcal{F}_{\beta\gamma}\mathcal{F}^{\beta\gamma}\\[4pt]
    &\= \tfrac12\sum_\alpha(\Theta'_\alpha)^2 - 
    2\bigl\{(\Theta_1{-}\Theta_2\Theta_0)^2+(\Theta_2{-}\Theta_0\Theta_1)^2+(\Theta_0{-}\Theta_1\Theta_2)^2\bigr\}\ .
\end{aligned} 
\end{equation}
This enjoys discrete symmetry of the permutation group $S_4$, acting by permuting $\{\Theta_\alpha\}$ and flipping the sign of any two $\Theta's$. Its maximal normal subgroups $S_3$ and $D_8$ yields {\it exact solutions}: 
\begin{itemize}
    \item {\bf Non-equivariant Abelian ans\"atze:} 
    \begin{equation}\label{abelAnsatz}
        \Theta_{0}\ \textrm{or}\ \Theta_{1}\ \textrm{or}\ \Theta_{2} \= \mathrm{h}(\psi) \quad\mathrm{while}\quad \textrm{rest} \= 0 \quad\implies\quad \mathrm{h}'' = -4\,\mathrm{h}\ ,
    \end{equation} 
    resulting into a harmonic equation whose solutions are well-known trigonometric functions.
    \item {\bf Equivariant non-Ablian ansatz:}
    \begin{equation}\label{nonAbelAnsatz}
        \Theta_{0}=\Theta_{1}=\Theta_{2}\ =:\ \tfrac{1}{2}\bigl(1+ \Phi(\psi) \bigr) \quad\implies\quad \Phi^{''}\ =\ 2\,\Phi\,(1-\Phi^{2})\ =\ -\sfrac{\pa V}{\pa \Phi}\ ,
    \end{equation}
    where the double-well potential $V(\Phi){=}\sfrac{1}{2}(\Phi^2-1)^2$ has been plotted in figure \ref{potential}. The resultant anharmonic equation here also admits analytic solutions in terms of Jacobi elliptic functions.
\end{itemize}

\section{$\textrm{AdS}_4$ gauge fields on Minkowski space}
We have already seen a series of conformal maps $(\tau,\rho,\psi) \rightarrow (\tau,\chi,\theta) \rightarrow (t,r,\theta)$ defined via \eqref{gluing} and \eqref{Mmap2} above. We can use these with $R{=}1$ along with abbreviations  $x\cdot \diff{x} := x_\mu\diff{x}^\mu$, $\varepsilon^1=\varepsilon^2:=\varepsilon$ and $\varepsilon^0:=1$ to write the $\textrm{AdS}_3$-cylinder one-forms $e^\alpha$ \eqref{1formsAdS} and $e^\psi:=\diff\psi$ in Minkowski coordinates as,
\begin{equation}\label{1formsMink}
\begin{aligned}
    e^{\alpha}&\= \sfrac{\varepsilon^\alpha}{\lambda^2{+}4z^2}\,\big(2(\lambda{+}2)\,\diff{x}^\alpha - 4x^\alpha\,x\cdot\diff{x} - 4\,f^\alpha_{\ \beta\gamma}\,x^\beta\diff{x}^\gamma \big)\ , \\
    e^{\psi}&\=\sfrac{\varepsilon}{\lambda^2{+}4z^2}\,\big({-}2\lambda\,\diff{z} + 4z\,x\cdot\diff{x} \big)\ ,\qquad \textrm{where}\quad \lambda\ :=\ r^2-t^2-1\ .
\end{aligned}
\end{equation} 
This allows one to read-off various vierbein components, viz.~$e^\alpha_\mu$ and $e^\psi_\mu$ to be used below.

\subsection{Nonabelian fields}
For the nonabelian case we use the the ansatz \eqref{nonAbelAnsatz} for gauge field $\Acal$ \eqref{gaugeField}. The corresponding field strength $F$ of $\mathcal{A}\equiv A = \tfrac{1}{2}\Big(1+\Phi\bigl(\psi(x)\bigr)\Big)\,I_{\alpha}\;e^{\alpha}_{\ \mu}\;\mathrm{d}x^{\mu}$ then computes to,
\begin{equation}
    F \= \tfrac{1}{2}\Big(\Phi'\bigl(\psi(x)\bigr)\,I_{\alpha}\,e^{\psi}_{\ \mu}e^{\alpha}_{\ \nu}\ -\ \tfrac{1}{2}\bigl(1{-}\Phi\bigl(\psi(x)\bigr)^{2}\bigr)\,I_{\alpha}\,f^{\alpha}_{\ \beta\gamma}\,e^{\beta}_{\ \mu}e^{\gamma}_{\ \nu} \Big)\,\diff x^\mu\wedge\diff x^\nu\ .
\end{equation}
One can easily extract the color-electromagnetic fields from this $F$ as follows: $E_{a}{:=}F_{a0}$ and $B_{a}{:=}\frac{1}{2}\epsilon_{abc}F_{bc}$. We can express the color EM fields thus obtained rather succinctly in terms of a Riemann--Silberstein vector ${\bf S}:={\bf E}+\im{\bf B}$ as,
\begin{align}
    S_{x} &= -\sfrac{2(\im\varepsilon\Phi'+\Phi^2{-}1)}{(\lambda{-}2\im z)(\lambda{+}2\im z)^2}\,
    \Bigl\{ 2\bigl[ty{+}\im x(z{+}\im)\bigr] I_0 + 2\varepsilon\bigl[xy{+}\im t(z{+}\im)\bigr] I_1 + \varepsilon\bigl[t^2{-}x^2{+}y^2{+}(z{+}\im)^2\bigr] I_2 \Bigr\}\ ,\\[5pt]
    S_{y} &= \sfrac{2(\im\varepsilon\Phi'+\Phi^2{-}1)}{(\lambda{-}2\im z)(\lambda{+}2\im z)^2}\,
    \Bigl\{ 2\bigl[tx{-}\im y(z{+}\im)\bigr] I_0 + \varepsilon\bigl[t^2{+}x^2{-}y^2{+}(z{+}\im)^2\bigr] I_1 + 2\varepsilon\bigl[xy{-}\im t(z{+}\im)\bigr] I_2 \Bigr\}\ ,\\[5pt]
    S_{z} &= \sfrac{2(\im\varepsilon\Phi'+\Phi^2{-}1)}{(\lambda{-}2\im z)(\lambda{+}2\im z)^2}\,
    \Bigl\{ \im\bigl[t^2{+}x^2{+}y^2{-}(z{+}\im)^2\bigr] I_0 + 2\varepsilon\bigl[\im tx{-}y(z{+}\im)\bigr] I_1 + 2\varepsilon\bigl[\im ty{+}x(z{+}\im)\bigr] I_2 \Bigr\}\ .
\end{align}
Interestingly, any explicit solution $\Phi$ do not couple to color components of these fields ${\bf S}$; this fact reflects below in that the corresponding physical quantities (arising from the stress-energy tensor) depends only on a conserved parameter---rather than an explicit form---of such solutions.

We can now compute the corresponding stress-energy tensor $T_{\mu\nu}$ given by,
\begin{align}\label{SEtensor}
    T_{\mu\nu}\=-\tfrac{1}{2g^{2}}\,\bigl( 
    \delta_\mu^{\ \rho}\delta_\nu^{\ \lambda}\eta^{\sigma\tau} -
    \tfrac14 \eta_{\mu\nu}\eta^{\rho\lambda}\eta^{\sigma\tau} \bigr)
    \,\mathrm{tr}\big(F_{\rho\sigma}F_{\lambda\tau}\bigr)\ ,
\end{align}
and express them, using mechanical energy $\epsilon:=-\tfrac14\bigl((\Phi')^2+(1{-}\Phi^2)^2\bigr)$, compactly as follows:
\begin{equation}
    \begin{split}
        \bigl(T_{\mu\nu}\bigr) \= \frac{8}{g^{2}}\,\frac{\epsilon}{(\lambda^2{+}4z^2)^3}
        \begin{pmatrix}
            \mathfrak{t}_{\alpha\beta} & \mathfrak{t}_{\alpha 3} \\
        \mathfrak{t}_{3\alpha} & \mathfrak{t}_{33}
        \end{pmatrix}
    \end{split}
    \with \left\{\
    \begin{split}
        \mathfrak{t}_{\alpha\beta} &\= -\eta_{\alpha\beta}(\lambda^2{+}4z^2)+16x_\alpha x_\beta z^2\,,\\
        \mathfrak{t}_{3\alpha} &\= \mathfrak{t}_{\alpha 3} \= -8x_\alpha z\,(\lambda{-}3z^2)\ ,\\
        \mathfrak{t}_{33} &\= 3\lambda^2-4z^2(1+4\lambda-4z^2)\ .
    \end{split}\right.
\end{equation} 
We find that these fields ${\bf E},{\bf B}$ and their stress-energy tensor $T_{\mu\nu}$ are not singular at the full hyperboloid $H^{1,2}\equiv\lambda{=}0$ but on a hypersurface given by the intersection
\begin{equation}\label{hypersurface}
    \{\lambda{=}0\}\cap\{ z{=}0\} \Leftrightarrow x^2+y^2-t^2 \= 1\ .
\end{equation}
We have plotted the energy density $\rho=\mathfrak{t}_{00}$ at $t{=}0$ highlighting the role of $xy$-plane in figure \ref{EnDenPlot}.

\subsection{Electromagnetic fields}
We consider following harmonic functions as solutions to the Abelian eom \eqref{abelAnsatz} such that they would have same structural form across the gluing surface (mediated by $\varepsilon$):
\begin{equation}
\begin{aligned}
    \widetilde{A}^{(0)}&\=-\tfrac12\,\cos2\bigl(\psi(x){+}\varepsilon\psi_{0}\bigr)\,e^{0}_{\ \mu}\;\mathrm{d}x^{\mu}\ ,\\[4pt]
    \widetilde{A}^{(1)}&\= \tfrac12\,\sin2\bigl(\psi(x){+}\varepsilon\psi_{0}\bigr)\,e^{1}_{\ \mu}\;\mathrm{d}x^{\mu}\ ,\\[4pt]
    \widetilde{A}^{(2)}&\= \tfrac12\,\sin2\bigl(\psi(x){+}\varepsilon\psi_{0}\bigr)\,e^{2}_{\ \mu}\;\mathrm{d}x^{\mu}\ .
\end{aligned}
\end{equation}
As before, we can go ahead and compute the corresponding field strengths $\widetilde{F}$, e.g.~this one
\begin{align}
    \widetilde{F}^{(0)}\= \bigl\{ \sin2\bigl(\psi(x){+}\varepsilon\psi_{0}\bigr)\,e^{\psi}_{\ \mu} e^0_{\ \nu}
    -\cos2\bigl(\psi(x){+}\varepsilon\psi_{0}\bigr)\,e^{1}_{\ \mu}e^{2}_{\ \nu}\big\}\;\mathrm{d}x^{\mu}\wedge \mathrm{d}x^{\nu}\ ,
\end{align}
and then extract the electric $\widetilde{{\bf E}}$ and magnetic $\widetilde{{\bf B}}$ fields. These can again be casted into a nice compact form using Riemann--Silberstein vector $\widetilde{{\bf S}}:=\widetilde{{\bf E}}+\im\widetilde{{\bf B}}$ as follows,
\begin{align}
    \widetilde{{\bf S}}^{(0)} &\= \frac{4\ep^{2\im\psi_0}}{(\lambda{+}2\im z)^3}
    \begin{pmatrix}
        -2\,\bigl(t\,y+\im x\,(z{+}\im)\bigr)\\[4pt]
        2\,\bigl(t\,x-\im y\,(z{+}\im)\bigr)\\[4pt]
        \im\,\bigl(t^2+x^2+y^2-(z{+}\im)^2\bigr)
    \end{pmatrix}\ ,\label{AbelRS0}\\[5pt]
    \widetilde{{\bf S}}^{(1)} &\= -\frac{4\ep^{2\im\psi_0}}{(\lambda{+}2\im z)^3}
    \begin{pmatrix}
        2\im\,\bigl(x\,y+\im t\,(z{+}\im)\bigr)\\[4pt]
        -\im\,\bigl(t^2+x^2-y^2+(z{+}\im)^2\bigr)\\[4pt]
        2\,\bigl(t\,x+\im y\,(z{+}\im)\bigr)
    \end{pmatrix}\ ,\label{AbelRS1}\\[5pt]
    \widetilde{{\bf S}}^{(2)} &\= \frac{4\ep^{2\im\psi_0}}{(\lambda{+}2\im z)^3}
    \begin{pmatrix}
        -\im\,\bigl(t^2-x^2+y^2+(z{+}\im)^2\bigr)\\[4pt]
        2\im\,\bigl(x\,y-\im t\,(z{+}\im)\bigr)\\[4pt]
        -2\,\bigl(t\,y-\im x\,(z{+}\im)\bigr)
    \end{pmatrix}\ .\label{AbelRS2}
\end{align}
Like before, these fields are also singular on the hypersurface \eqref{hypersurface}. We demonstrate this in figure \ref{fieldLines} by plotting typical field lines for $\widetilde{{\bf E}}^{(0)}$ and $\widetilde{{\bf B}}^{(0)}$ \eqref{AbelRS0}; these accumulate and intersect at the singular boundary (denoted with black circle). We omit here the expressions for the stress-energy components (due to their bulky form) and refer the reader to \cite{Hir23}, where such expressions corresponding to fields in \eqref{AbelRS0} have been noted.

Incidentally, the above Riemann--Silberstein vectors are reminiscent of the Hopf--Ran{\~a}da (HR) electromagnetic knots \cite{zhilin} and suggests their interpretation as a non-compact cousin of the HR knot. Upon further exploration we find that a special case of the field configuration \eqref{AbelRS0}:
\begin{equation}\label{AbelAdS}
    \psi_0=z=0 \quad\implies\quad
    \frac{4}{(x^2{+}y^2{-}t^2{-}1)^3}\begin{pmatrix}
        2(x-t\,y) \\[4pt] 2(y+t\,x) \\[4pt] t^2{+}x^2{+}y^2{+}1
    \end{pmatrix}\ ,
\end{equation}
reproduces the magnetic fields\footnote{In our convention, these gauge fields are interpreted as two electric fields and one magnetic field.} of a recently constructed magnetic vortex~\cite[eq. (5.15)]{RS18} under the following identification of the $\textrm{AdS}_3$ coordinates $(x^0,x^1,x^2)$ used in \cite{RS18},
\begin{equation}
    x^0 \= -t\ ,\quad x^1 \= -y \und x^2 \= x\ .
\end{equation}

\section{Summary}
\begin{itemize}
    \item First, we employed the $\text{AdS}_3$-slicing of $\text{AdS}_4$ and the group manifold structure of $\text{SU}(1,1)$ to find Yang--Mills solutions on $\Ical\times \text{AdS}_3$.
    \item Next, we used a series of conformal maps to transfer these solutions to Minkoowsi space, since Yang--Mills theory is conformally invariant in $4$-dimensions.
    \item These gauge fields are singular on a $2$-dimensional hyperboloid $x^2{+}y^2{-}t^2=1$, but this singularity is milder than the one we started with, namely $r^2{-}t^2{-}1=0$.
    \item Due to this singularity the total energy diverges for both kinds of gauge fields (see \cite{Hir23} for details), thereby limiting their physical usefulness.
    \item Nevertheless, our Abelian solutions were found to match the magnetic field of a known vortex magnetic mode on SU$(1,1)$ \cite{RS18}. The status of other two Abelian solutions in this regard remains to be explored.
\end{itemize}

\end{document}